\documentclass[longauth]{aa} % for the long lists of affiliations 
\usepackage{graphicx}
\usepackage{txfonts}
\begin{document}
   \title{Transit timing analysis of CoRoT-1b
         %No Neptune mass companion in the short 
         %period range
	 \thanks{Based on observations obtained with CoRoT, a space 
         project operated by the French Space Agency, CNES, 
         with participation of the Science Programs of 
         ESA, ESTEC/RSSD, Austria, Belgium, Brazil, Germany, 
         and Spain}}

   \author{    Csizmadia, Sz.\inst{1}
        \and S.~Renner \inst{2,3}
        \and P.~Barge \inst{4}
  	\and E.~Agol \inst{23}
  	\and S.~Aigrain \inst{8}
        \and R.~Alonso \inst{4,20}
  	\and J.-M.~Almenara \inst{7}
        \and A.~S.~Bonomo \inst{4, 22}
  	\and P.~Bord\'e \inst{9}
  	\and F.~Bouchy \inst{10}
        \and J.~Cabrera \inst{1, 5}
  	\and H.~J.~Deeg \inst{7}
  	\and R.~De la Reza  \inst{11}
  	\and M.~Deleuil \inst{4}
  	\and R.~Dvorak \inst{12}
        \and A.~Erikson \inst{1}
        \and E.~W.~Guenther \inst{7,15}
  	\and M.~Fridlund  \inst{13}
  	\and P.~Gondoin \inst{10}
  	\and T.~Guillot  \inst{14}
  	\and A.~Hatzes \inst{15}
  	\and L.~Jorda  \inst{4}
  	\and H.~Lammer  \inst{16}
  	\and C.~L\'azaro \inst{10,17}
  	\and A.~L\'eger  \inst{9}
  	\and A.~Llebaria \inst{4}
  	\and P.~Magain \inst{18}
  	\and C.~Moutou  \inst{4}
  	\and M.~Ollivier  \inst{9}
  	\and M.~P\"atzold  \inst{19}
  	\and D.~Queloz  \inst{20}
        \and H.~Rauer \inst{1, 6}
  	\and D.~Rouan  \inst{21}
  	\and J.~Schneider  \inst{5} 
  	\and G.~Wuchterl \inst{15}
  	\and D.~Gandolfi \inst{15}
}

\small{
\offprints{szilard.csizmadia@dlr.de (Sz. Csizmadia)}

   \institute{%1
	      Institute of Planetary Research, DLR, Rutherfordstr. 2, 
              12489 Berlin, Germany
              \email{szilard.csizmadia@dlr.de}
              \and % 2
              Institut de M\'ecanique C\'eleste et de Calcul de Eph\'em\'erides, 
	      UMR 8028 du CNRS, 77 avenue Denfert-Rochereau, 75014 Paris, France
	      \and % 3
              Laboratoire d'Astronomie de Lille, 
	      Universit\'e Lille 1, 1 impasse de l'observatoire, 59000 Lille, France
	      \and % 4
	      Laboratoire d'Astrophysique de Marseille, 
	      UMR 6110, CNRS/Universit\'e de Provence, 38 rue F. Joliot-Curie, 13388
              Marseille, France
	      \and % 5
	      LUTH, Observatoire de Paris, CNRS, Universit\'e Paris Diderot; 5
	      place Jules Janssen, 92190 Meudon, France 
	      \and % 6
	      Center for Astronomy and Astrophysics, TU Berlin, 
	      Hardenbergstr. 36, D-10623 Berlin, Germany
	      \and %  7
	      Instituto de Astrof\'isica de Canarias, E-38205 La Laguna, Tenerife,
  	      Spain
              \and %  8
              School of Physics, University of Exeter, Stocker Road, Exeter EX4
              4QL, United Kingdom
              \and %  9
              Institut d'Astrophysique Spatiale, Universit\'e Paris XI, F-91405
              Orsay, France
              \and %  10
              Institut d'Astrophysique de Paris, Universit\'e Pierre \& Marie
              Curie, 98bis Bd Arago, 75014 Paris, France
  	      \and %  11
              Observat\'orio Nacional, Rio de Janeiro, RJ, Brazil
              \and %  12
              University of Vienna, Institute of Astronomy, T\"urkenschanzstr. 17,
              A-1180 Vienna, Austria
              \and % 13
              Research and Scientific Support Department, ESTEC/ESA, PO Box 299,
              2200 AG Noordwijk, The Netherlands
  \and % 14
  Observatoire de la C\^ote d\'Azur, Laboratoire Cassiop\'ee, BP 4229,
  06304 Nice Cedex 4, France
  \and % 15
  Th{\"u}ringer Landessternwarte, Sternwarte 5, Tautenburg 5, D-07778
  Tautenburg, Germany
  \and % 16
  Space Research Institute, Austrian Academy of Science,
  Schmiedlstr. 6, A-8042 Graz, Austria
  \and % 17
  Departamento de Astrof\'isica, Universidad de La Laguna, E-38205 La
  Laguna, Tenerife, Spain
  \and % 18
  University of Li\`ege, All\'ee du 6 ao\^ut 17, Sart Tilman, Li\`ege
  1, Belgium
  \and % 19
  Rheinisches Institut f\"ur Umweltforschung an der Universit\"at  zu
  K\"oln, Aachener Strasse 209, 50931, Germany
  \and % 20
  Observatoire de Gen\`eve, Universit\'e de Gen\`eve, 51 chemin des
  Maillettes, 1290 Sauverny, Switzerland
  \and % 21
  %ESIA, Observatoire de Paris-Meudon, 5 place Jules Janssen, 92195
  %Meudon, France
  LESIA, UMR 8109 CNRS , Observatoire de Paris, UVSQ, Universit\'e
  Paris-Diderot, 5 place J. Janssen, 92195 Meudon, France
  \and % 22
  INAF - Osservatorio Astrofisico di Catania, via S. Sofia 78, 
  95123 Catania, Italy
  \and % 23
  Department of Astronomy, University of Washington, Box 351580, Seattle, 
  WA 98195, USA
  }

          }

   \date{Received 13 March 2009 / Accepted 2 November 2009}

% \abstract{}{}{}{}{} 
% 5 {} token are mandatory
 
  \abstract
   {CoRoT, the pioneer space-based transit search, steadily provides thousands
   of high-precision 
   light curves with continuous time sampling over periods of up to 5 months. The transits of a 
   planet perturbed by an additional object are not strictly periodic. By studying the transit timing
   variations (TTVs), additional objects can be detected in the system.}
   {A transit timing analysis of CoRoT-1b is carried out to constrain the existence of additional planets
   in the system.
   %We carried out a transit timing analysis of CoRoT-1b to search for possible orbital 
   %period variations. {\bf Several dynamical models were calculated to determine
   %the maximum possible values of a hypothetical perturber object.}
   }
   { We used data obtained by an improved version of the CoRoT data 
    pipeline (version 2.0). Individual transits were fitted to determine the 
    mid-transit times, and we analyzed the derived $O-C$ diagram. N-body
    integrations were used to place limits on secondary planets.}
   {No periodic timing variations with a period shorter 
   than the observational window (55 days) are found. The presence of an 
   Earth-mass Trojan 
   is not likely. A planet of mass greater
   than $\sim 1$ Earth mass can be ruled out by the present data if the object
   is in a 2:1 (exterior) mean motion resonance with CoRoT-1b. Considering
   initially circular orbits: (i) super-Earths (less than 10 Earth-masses) are
   excluded for periods less than about 3.5 days, (ii) Saturn-like planets can
   be ruled out for periods less than about 5 days, (iii) Jupiter-like planets
   should have a minimum orbital period of about 6.5 days.}
   {}

   
   \titlerunning{TTVs in CoRoT-1b}

   \maketitle
%
%________________________________________________________________

\section{Introduction}

   %\cite{mizuno})

As a consequence of the gravitational perturbations, the mid-times of
consecutive transits deviate from a linear ephemeris in a transiting
exoplanet system (transit timing variation, hereafter TTV). Depending on
the mass and the orbital configuration of the perturbing planet, this
deviation can have amplitudes from a few seconds to days (e.g. Steffen, 2006). 
Moreover, the
duration, shape, and depth of the transits can also change. In extreme
cases the transits can disappear and then reappear (Schneider 1994, 2004).
Both the theoretical aspects and the observable effects have been studied in
e.g. Miralda-Escud\'e (2002), Borkovits et al. (2003), Agol et al.
(2005), Holman \& Murray (2005), Ford \& Holman (2007), Simon et al.
(2007), Heyl \& Gladman (2007), Steffen et al. (2007), Nesvorn\'y \&
Morbidelli (2008), P\'al \& Kocsis (2008) and Kipping (2009). Several
transiting exoplanets were subject to this kind of analysis (Steffen \& Agol
2005, Agol \& Steffen 2007, Miller-Ricci et al. 2007, Alonso et al.
2008, Hrudkov\'a et al. 2008, Miller-Ricci et al. 2008ab, Diaz et al. 2008,
Coughlin et al. 2008, Rabus et al. 2009, Stringfellow et al. 2009).

In addition, stellar spots may affect the transit shape and, because
of this, we have some difficulty in determining of the midtime of
the transit. This effect may cause spurious periodic terms in the $O-C$
diagram of exoplanets (Alonso et al. 2008, Pont et al. 2007).

Here we report the TTV analysis of CoRoT-1b based on data obtained by an
improved version of the CoRoT data pipeline. In this system a low-density
planet (mass: 1.03$M_{Jup}$, radius: 1.49$R_{Jup}$, average density:
$0.38~gcm^{-3}$, semi-major axis: 5.46$R_{\odot}$, orbital period:  $1.5089557$
days) orbits a G2 main-sequence star (Barge et al. 2008). Thirty-six transits
were observed by CoRoT, 20 of them in 512 seconds and 16 with the 32 seconds
sampling rate mode. In total more than 68 000 data points were collected during
55 consecutive days (Barge et al. 2008). The operation of the satellite is
described in detail in the pre-launch book, and the reader can find useful
information about CoRoT in Baglin et al. (2007), Boisnard et al. (2006), and
Auvergne et al. (2009).

\section{Methods of mid-transit point determination: effect of the sampling rate
on the precision}

If one uses the CoRoT data for TTV analysis, the main limiting factor
arises from the sampling rate. The typical length of the ingress/egress
phase of a hot Jupiter is on the order of 10-20 minutes.  (In the
particular case of CoRoT-1b, the ingress/egress time is 9.8 minutes.)
CoRoT targets are observed with 512 or 32 second sampling rates (the
so-called undersampled/oversampled modes, see Fig~1). Concerning a
typical 3 hour transit, one can easily conclude that a transit
observation consists of only $(3 \times 3600 \mathrm{seconds}) / 512
\mathrm{seconds} \approx 21$ data points. In the oversampled mode
we have typically over 300 data points per transit. The small number of
data points in the undersampled mode may not be balanced by the very
good photometric precision of CoRoT (which is about 0.1\% for a 13
magnitude star in white light for a 512 second exposure, see Auvergne
et al. 2009), therefore we chose to investigate this issue.

A second factor arises from the orbit of the satellite. The satellite
periodically crosses the so-called South Atlantic Anomaly (SAA) region, which
causes bad/uncertain data points and long data gaps (typically 10 minutes).
This is significant only when the SAA-crossing occurs during the ingress or the
egress phase. Therefore the following test was carried out. Using the exoplanet
light curve model of Mandel \& Agol (2002) and the parameters of the system
(Barge et al., 2008), we simulated the light curve of CoRoT-1b. Then this curve
was re-sampled to the same time-points as CoRoT observations. We added a
Gaussian-like random noise to the points. The standard deviation of the noise
term was chosen in such a way that we had the same signal-to-noise ratio as
given in Barge et al. (2008) for the CoRoT-1 light curve. A constant orbital
period was assumed.

%                                     Two column figure
%______________________________________________ 
   \begin{figure*}
   \centering
   \includegraphics[width=6.9cm]{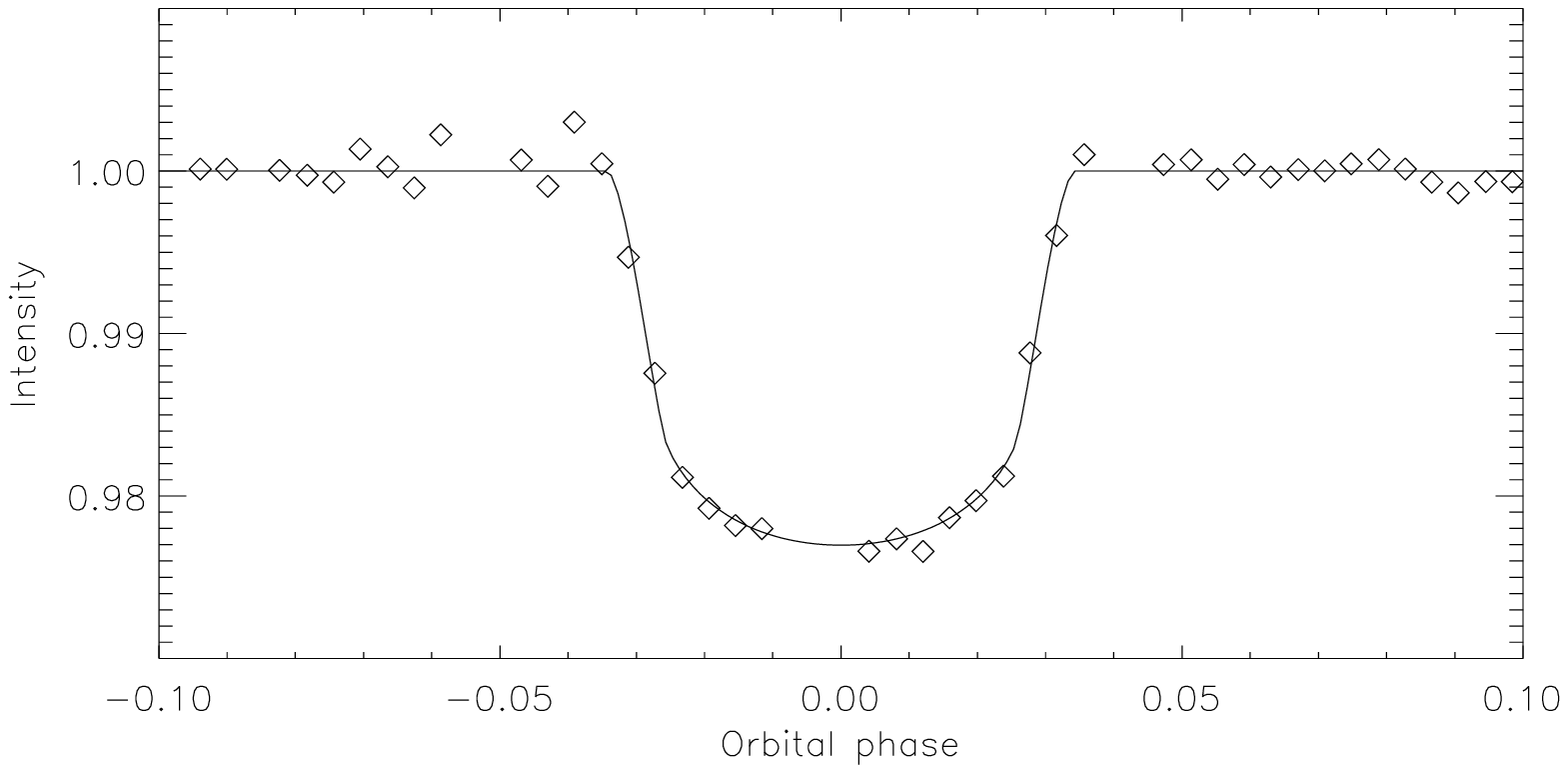}
   \includegraphics[width=6.9cm]{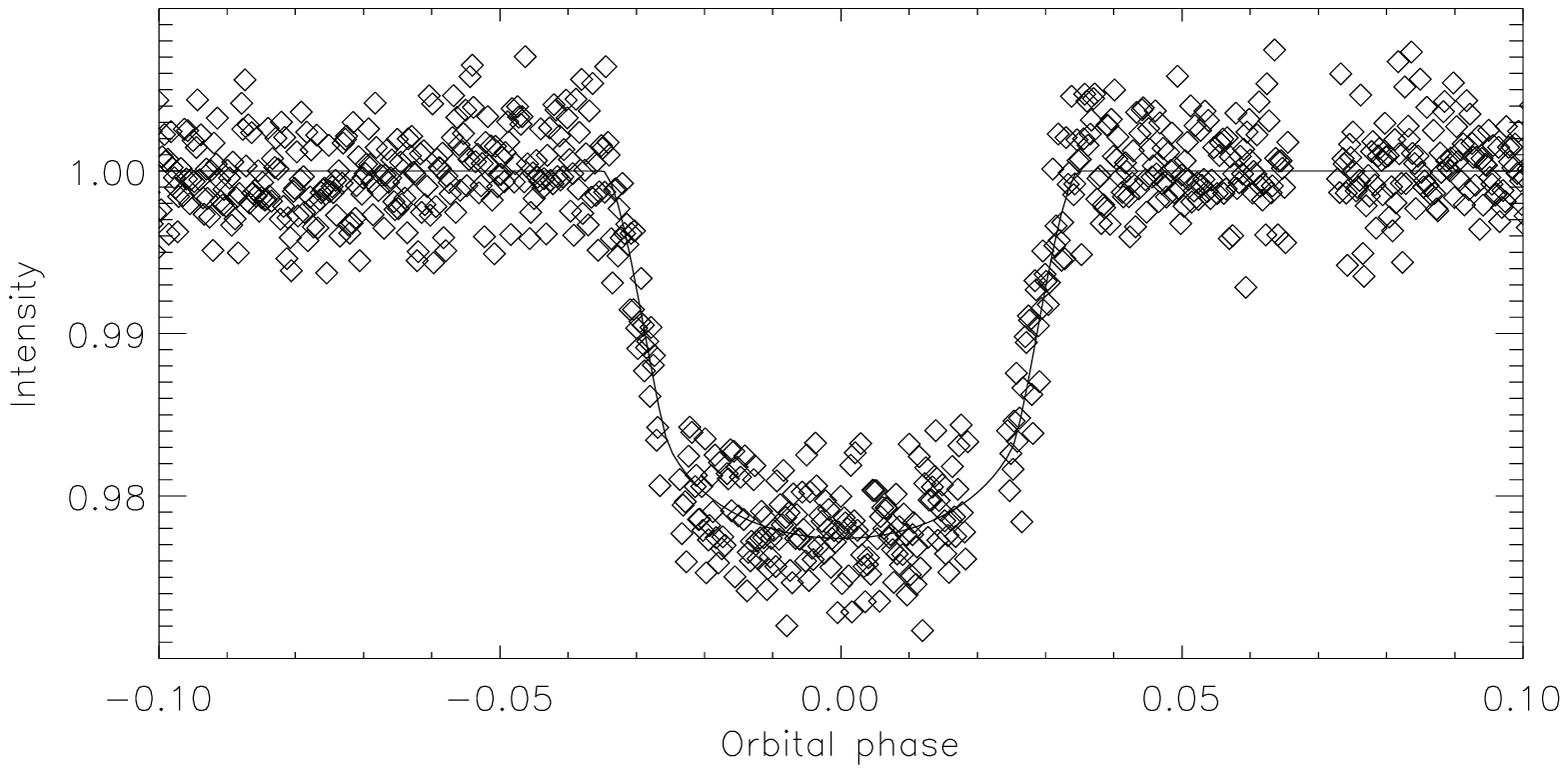}
   \caption{Top: A typical transit observation of CoRoT-1b by CoRoT in
the 512 sec sampling rate mode. Bottom: A typical transit observation of CoRoT-1b by CoRoT 
in the 32 sec sampling rate mode. Abcissa is phase, ordinate is 
normalized intensity. The solid lines show the fit.}
              \label{Fig12}%
    \end{figure*}

Then we determined the mid-transit times in this simulated light curve,
fitting each individual transit separately. Again, we use the
Mandel \& Agol (2002) model combined with the Amoeba algorithm (Press et al.
1992) to find the optimum fit. We assumed that the planet-to-stellar
radius ratio and the limb-darkening coefficients are known, so they were fixed. 
%Again we used a fixed orbital period. 
The adjustable parameters are the mid-transit point, the inclination, and the 
$a/R_s$ ratio ($a$: semi-major axis, $R_s$: stellar radius). 

On average, the fits of the individual transits yield only a difference
of 9 seconds between the real and the determined midtransit points in
the undersampled mode, when there are data points in the ingress/egress
part of the light curve. When the ingress or egress part is missing in
the undersampled mode, the errors can be as large as 20-60 seconds,
depending on the distribution of points during the transit. If both the
ingress and egress parts are missing, the errors are 60-120 seconds,
sometimes even more.

These optimistic error bars should be increased due to at least two
different effects. First, we do not know a priori the exact value of the
orbital period which leads to small uncertainties in the calculation of
the phase. We estimate that for CoRoT-1b this is negligible. Second, the
stellar activity is not included. However, the detailed discussion of
these two effects goes beyond the purpose of the present investigation.

Since we use a constant period and assume e=0 during the simulation, we
expect a linear $O-C$ curve with some scatter. To better
characterize this scatter we calculate the standard deviation of the
sample. We find that the mean $1\sigma$ scatter of the resulting overall
$O-C$ diagram of this simulated light curve with a constant period is 22
seconds. But it is 27 seconds for the undersampled part and 16 seconds
for the oversampled part.

%We repeated this analysis with the assumption that the orbital period can vary,
%but all planet parameters as well as inclination and semi-major axis are
%constant. Then the O-C diagram is better characterized: the $O-C$ data points
%show 27 seconds $1\sigma$ scatter, while it is 37 seconds for the undersampled
%part and is 21 seconds for the oversampled part, respectively. Therefore if we 
%can assume that the shape of the transit light curve does not change in time, 
%more precise results can be derived by fixing all the parameters except the
%centre of the transit. However, the validity of such an assumption is limited
%since it requires that the size, eccentricity and the orientation of the orbit
%cannot be changed which is not the case in general.

\section{TTV analysis of CoRoT-1b}

We used the N2 level data points (Auvergne et al. 2009) processed by the 
2.0 version of the pipeline
(not yet realeased data for the public). The resulting light curve was manually
checked: a few data points were noted by the pipeline to be affected by 
cosmic ray events in spite of it having no problems -- we 
restored these data points. In addition, several outliers were removed by hand.
%(From time to time the pipeline regards certain
%points as a cosmic ray event in spite of the fact that they are not affected. We
%corrected this by visual inspection of the light curve. 
%In addition, several bad data points were removed by hand during the visual inspection of 
%the individual transits.) 
Then we performed a TTV analysis by fitting all transits using the method described in
the previous section. Transit No. 30 is excluded from this investigation
because it is strongly affected by noise. Table~1 gives the midtransit times 
and their errors.

%                                     Two column figure
%______________________________________________ 
   \begin{figure*}
   \centering
   \includegraphics[width=6.9cm]{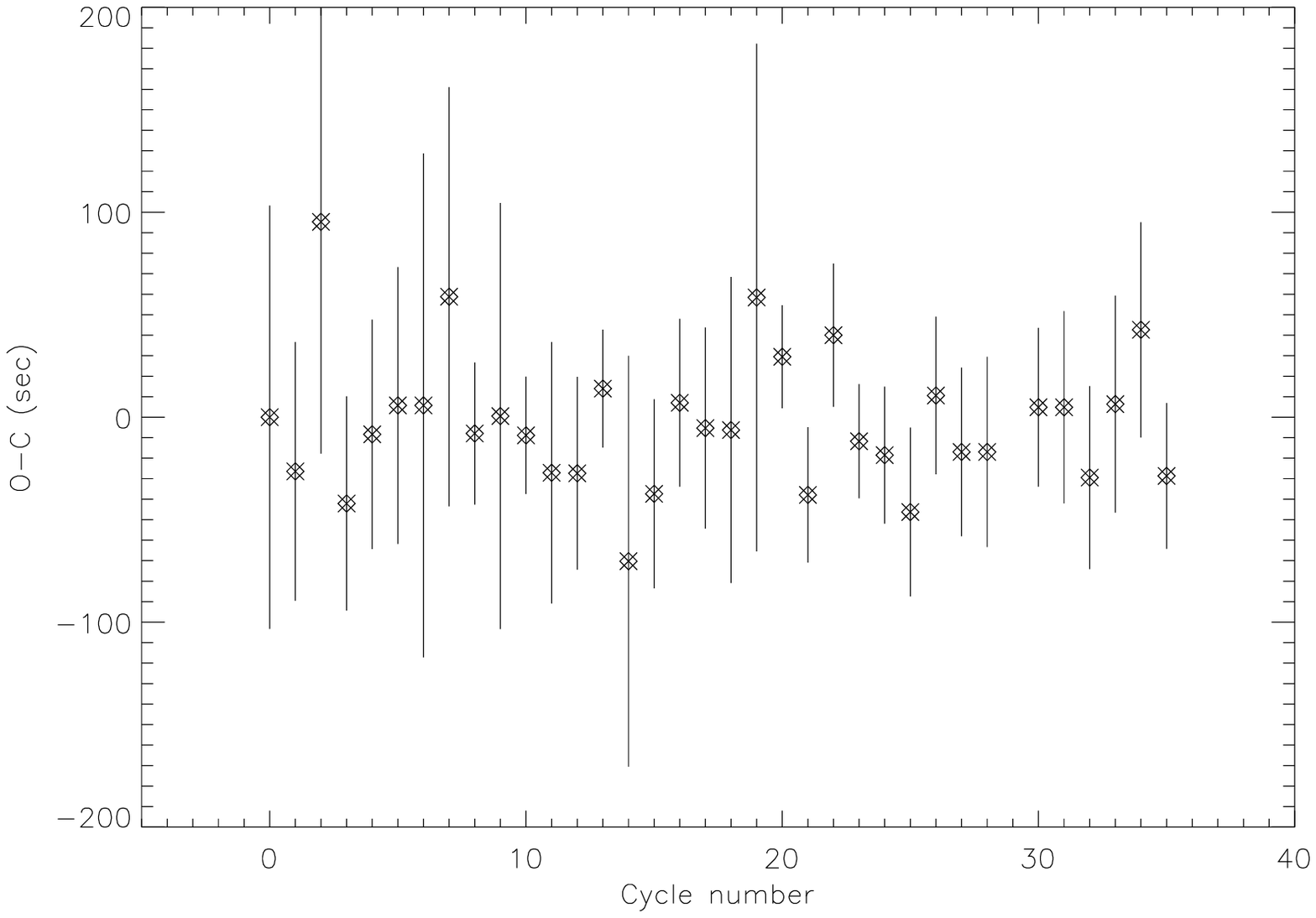}
   \includegraphics[width=6.9cm]{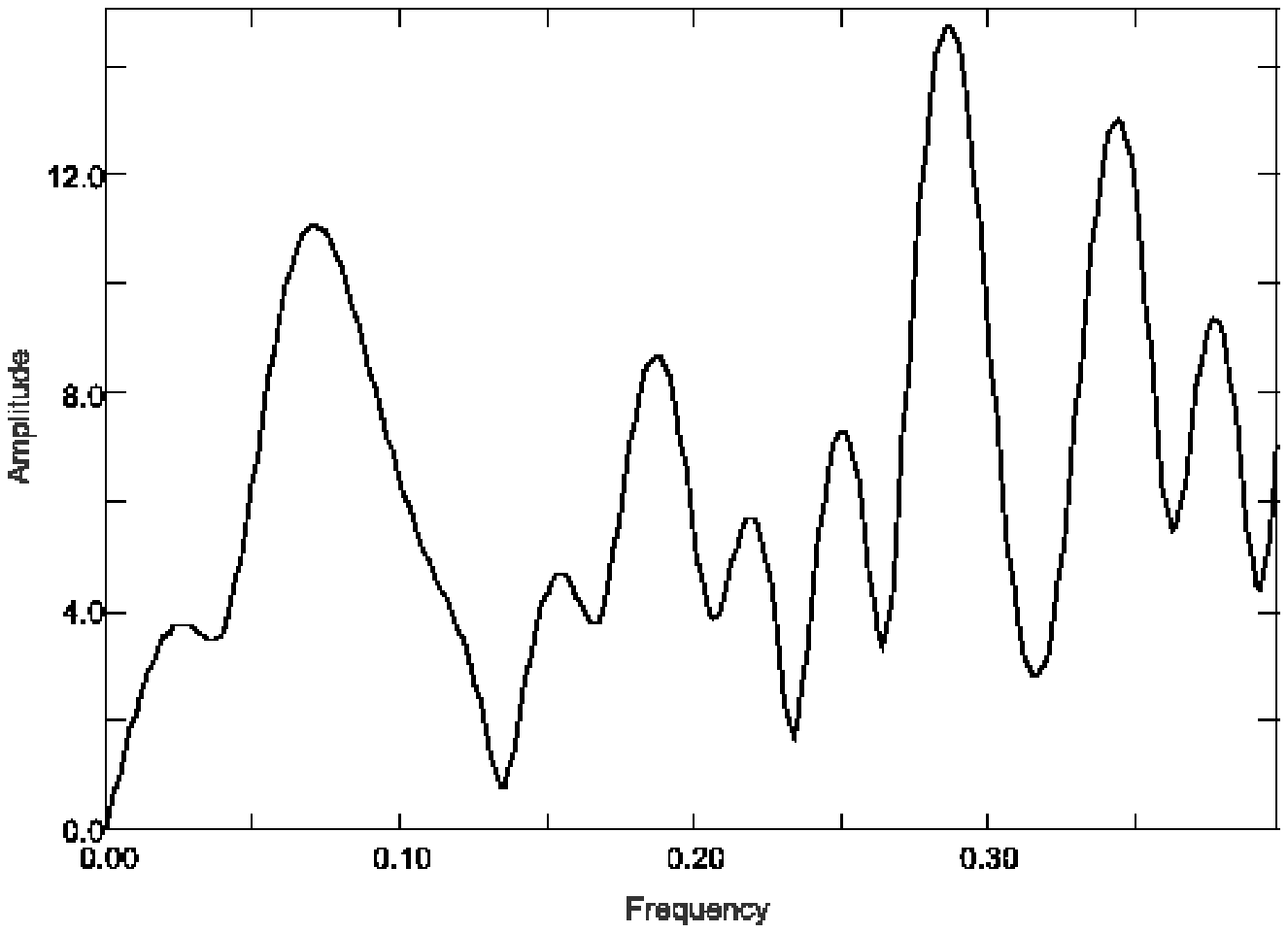}
   \caption{Top: The $O-C$ diagram of CoRoT-1b obtained by the light curve fit of the
            individual transits. Bottom: The power spectrum of the
	    Fourier-transform of the $O-C$ diagram. Frequency is in cycles/day and amplitude is in
	    seconds.}
              \label{Fig34}%
    \end{figure*}
%

%\begin{figure*}%f3
%\includegraphics[width=6.9cm]{1b_oc.eps}
%\caption{The $O-C$ diagram of CoRoT-1b obtained by light curve fit of the
%individual transits.}
%\label{fig7}
%\end{figure*}
%
%\begin{figure*}%f4
%\includegraphics[width=6.9cm]{1b_TTV1.eps}
%\caption{The power spectrum of the Fourier-transform of the $O-C$ diagram.}
%\label{fig7}
%\end{figure*}

%__________________________________________________ 
   \begin{table}
      \caption[]{Mid-transit times (HJD) and errors (days) of CoRoT-1b.}
         \label{oc1}
     	\begin{tabular}{cccc}
            \hline
            \noalign{\smallskip}
            HJD & Error (days) & HJD & Error (days)\\
            \noalign{\smallskip}
            \hline
            \noalign{\smallskip}
2454138.32782 & 0.00039 & 2454165.48897 & 0.00028 \\
2454139.83657 & 0.00024 & 2454166.99842 & 0.00047 \\
2454141.34646 & 0.00043 & 2454168.50716 & 0.00019 \\
2454142.85436 & 0.00020 & 2454170.01559 & 0.00025 \\
2454144.36357 & 0.00021 & 2454171.52515 & 0.00026 \\
2454145.87264 & 0.00025 & 2454173.03371 & 0.00021 \\
2454147.38159 & 0.00047 & 2454174.54261 & 0.00025 \\
2454148.89096 & 0.00039 & 2454176.05135 & 0.00031 \\
2454150.39940 & 0.00013 & 2454177.56074 & 0.00029 \\
2454151.90842 & 0.00039 & 2454179.06949 & 0.00031 \\
2454153.41730 & 0.00010 & 2454180.57845 & 0.00035 \\
2454154.92612 & 0.00024 & 2454183.59652 & 0.00029 \\
2454156.43507 & 0.00018 & 2454185.10548 & 0.00035 \\
2454157.94435 & 0.00011 & 2454186.61417 & 0.00034 \\
2454159.45266 & 0.00038 & 2454188.12340 & 0.00040 \\
2454160.96186 & 0.00017 & 2454189.63264 & 0.00040 \\
2454162.47116 & 0.00015 & 2454191.14105 & 0.00027 \\
2454163.98002 & 0.00018 &               &         \\
            \noalign{\smallskip}
            \hline
\end{tabular}
\end{table}

The overall $O-C$ diagram and its Fourier-transform are given in Fig. 2. 
This diagram is built using the observed light curve.  It is prominent
that after switching on the 32 sec sampling rate mode (after the 20th transit),
the $1\sigma$ scatter of the $O-C$ diagram is reduced to only 18 seconds, compared to
the $1\sigma$ scatter of the 25 seconds observed in the undersampled mode.  The
$1\sigma$ scatter of the whole $O-C$ diagram is 22 seconds. All these scatter
values are very close to the value we would expect in the case of a constant
orbital period (see previous section).

No clear periodicity or trend can be identified in this diagram. We calculate
the Fourier-spectrum of the $O-C$ curve by the Period04 software (Lenz \& Breger 2005) to
search for any non-obvious periodicity. The power spectrum shows few peaks, 
but none of them is above the noise level. The highest peak has only $S/N
\approx 2$, so it is not regarded as a real signal.

We conclude that there are not any periodic TTVs in CoRoT-1b with a period
less than the observational window (55 days) and an amplitude larger than
about 1 minute (=$3\sigma$ detection level). We also note that there is no
significant change in the inclination and the $a/R_{star}$ ratio during this
interval. Bean (2009) presents results about the TTV analysis of CoRoT-1b
based on the public data processed by an earlier version of the
pipeline. His O-C diagram shows a larger $1\sigma$-scatter (36 seconds
compared to our 22 seconds).

\section{Limits on secondary planets}

\subsection{ Examples of simulated TTVs}

We show in Table~2 which amplitudes and periods can typically be 
expected in the $O-C$ diagram of the CoRoT-1b system based on some
dynamical simulations. We include an Earth-mass planet at the $L_4$
Lagrangian point, an Earth in the 2:1 exterior mean motion resonance, a
nearby Neptune-like planet, or an outer eccentric Neptune. The stellar
mass, the mass of CoRoT-1b and its orbital elements are fixed to the
values given in Barge et al. (2008). We use the Mercury software
(Chambers 1999), with the Burlish-Stoer algorithm (accuracy parameter
$\delta=10^{-16}$). As one can see {in Table~2}, $O-C$ variations on the
order of 60-150 seconds might occur on a short time scale (typically
10-150 orbital cycles of the transiting planet).

{\it Case 1}: an Earth-mass Trojan planet librating with an amplitude of $20^\circ$ around
the $L_4$ Lagrangian point would have an amplitude of 60 seconds in the $O-C$ diagram with a
period of about 10 orbital cycles of the transiting planet (about 15 days, see Table~2). 
The amplitude is close to our detection limit. There is a peak in
the Fourier-spectrum of the $O-C$ diagram at the corresponding frequency with an amplitude 
of about 11 seconds. However, the peak is not significant (S/N = 1.3 only). Therefore,
an additional planet with similar parameters is not likely.

{\it Case 2}: An Earth-mass planet initially on a circular orbit and in 2:1 mean
motion resonance with CoRoT-1b would have an amplitude of about 100 seconds in the 
$O-C$ diagram with a period of about 150 transits (approximately 225 days, see Table 2 
and Fig.~3). 
If the CoRoT observational window was around the maximum or the
minimum of the $O-C$ curve (see Fig. 3) then we would have no chance to discover 
this possible planet because the amplitude is on the order of the scatter. 
%If the observational window matched the part between the maximum and minimum,
%the variation of the $O-C$ values during 36 transits (our observational window)
%could be as large as 80 seconds, larger than our detection limit. Since we do
%not know which part of such a diagram was observed, we cannot exclude the
%possibility of such a planet. 
If the observational window matched the steepest part of the $O-C$ diagram, 
we would observe a linear $O-C$ curve that could be interpreted as a
wrong period value. This gives a hint: if there are no observed period
variations in a short observational window, this does not mean that we can give 
an upper limit for a hypothetical perturber object. It might be the case that 
we are on a linear part of the $O-C$ curve. The observational window should be 
long enough to exclude similar cases.

{\it Cases 3 and 4}: simulations show that an outer 30 Earth-mass
planet, close to CoRoT-1b ($P=2.772118632$ days and $e=0.05$) or eccentric
($P=4.2679123$ days and $e=0.25$), cause $O-C$
variations of about 150 seconds, within approximately 15 and 30
orbital revolutions of the transiting planet, respectively (see Table 2). 
This is much greater than our detection limit, so outer planets in the CoRoT-1b system
with similar orbital parameters can be excluded.

%{\bf In the next subsection we investigate the possibility of the presence of a
%second object in the system detailed.}

\subsection{Detailed analysis}

Using N-body integrations, we computed the maximum mass of a hypothetical
perturbing planet, with given initial orbital periods and eccentricities,
leading to TTVs compatible with the data. To calculate the transit times, we
used a bracketing routine from Agol et al. (2005). The orbits of CoRoT-1b and an
additional planet were computed over the timespan of the observations, using a
Burlish-Stoer integrator with an accuracy parameter $\delta=10^{-16}$. The
equations of motion were integrated in a cartesian reference frame centered on
the barycenter of the system. The transit times are subtracted from the data to
give the O-C residuals and $\chi^2$.

The masses of the central star and
CoRoT-1b are respectively fixed at 0.95 solar masses and 1.03 Jupiter masses
(Barge et al. 2008). The orbit of CoRoT-1b is initially circular with an orbital
period $P=1.5089557$ d (Barge et al. 2008) and a true longitude 
$\theta=0$ deg. With these parameters and without any perturbation due to an additional body,
the first transit occurs at $T(HJD)=2454138.327840$, and the residuals given by
the numerical integration are at their minimum (i.e. the same as the ones from the best constant
period fit, see Sect. 3) with the following values: zero mean, standard deviation
$\sigma_{min}=21.62$ seconds, $\chi^2=24.55$. 

The perturbing planet is assumed to lie on the same plane as CoRoT-1b.
For given initial orbital parameters, we increase the mass of the test
planet, starting at 0.1 Earth masses, and calculate the standard
deviation of the O-C residuals. We store the mass value for which this
rms exceeds the observed scatter $\sigma_{min}$. In this way we
determine the maximum planet's mass allowed. The results are shown in
Fig. 4 (respectively Fig. 5), which shows the maximum mass for a
perturbing object as a function of its initial orbital period (resp.
initial orbital period and eccentricity). In Fig. 4, the mass of the
secondary planet has been varied between 0.1 and 100 Earth masses (100
values on a log scale), and its initial orbital period between 2.8 and
7.6 days (with a step of 0.0015 days). For any given orbital
period, eccentricity, and mass value, the TTV-signal is computed over
the range of possible initial true anomaly and longitude of pericenter
values to minimize the resulting residuals. In Fig. 5, the perturbing
planet is initially at its apocenter (fixed at 180 degrees from
CoRoT-1b), and the following grid of parameters has been used: (i)
masses between 0.1 and 100 Earth masses (100 values on a log scale),
(ii) orbital periods between 2.8 and 7.6 days (with a step of 0.001333
days), (iii) eccentricities between 0 and 0.25 (100 values on a log
scale).

From Fig. 4, Saturn-like planets can be ruled out for periods less
than about 5 days if e=0 (respectively 6 days if e=0.2). As shown in the
figures, perturbing planets with eccentric orbits obviously cause larger
TTVs, hence have lower mass limits. Super-Earths are defined as
planets with 1-10 Earth masses (Valencia et al. 2007). Depending on the
initial eccentricity, such planets with orbital periods less than
3.4-4.1 days can be excluded. Planets with masses greater than 0.3-1.0
Earth masses can be ruled out by the data if they are in the 2:1
(exterior) mean motion resonance with CoRoT-1b.  The data do
not allow strongly constraining the mass of perturbing planets near
higher order resonances. Finally, we estimate the minimum orbital
period for an outer Jupiter-mass planet. From Holman and Murray (2005),
\begin{equation}
M_\mathrm{perturber} = \frac{16\pi}{45} M_\mathrm{star} 
\frac{\Delta t_{max}}{P_\mathrm{transiting}} 
\left(\frac{P_\mathrm{perturber}}{P_{transiting}} \right )^2 
(1-e_\mathrm{perturber})^3 
\end{equation} 
When assuming a circular orbit and $\Delta t_{max} = 3\sigma_{min}$, this
yields a minimum orbital period of $2.0$ days. Otherwise, we would see
its effect in the $O-C$ diagram. This lower limit is in good agreement with the
numerical simulations (see Figs. 4 and 5)

\section{Summary}

Our work shows that CoRoT allows study of the short time scale 
(30 days for the Short Run fields, 150 days for the Long Run fields)
transit timing variations whose $1\sigma$ detection limit depends on the
sampling rate, and it is 22 seconds for CoRoT-1b. The comparison of the
$O-C$ diagram of CoRoT-1b with numerical integrations leads to the
following results: (i) an Earth-mass planet at the $L_4$ point is not
likely. If existing, its detectability would be close to the $3\sigma$
detection limit, (ii) an outer Earth-mass planet in 2:1 resonance with
CoRoT-1b can be rejected, given our data set.  However, a longer
observational window is required to fully assess the presence of such a
planet, (iii) super-Earths are excluded for periods less than about 3.5
days, (iv) Saturn-like planets are ruled out for periods less than about
5 days.

Bean (2009) finds that there is no additional planet in the system with 4
Earth-mass or greater on an orbit with 2:1 mean motion resonance. Using an
improved version of the CoRoT data pipeline, we confirm his result.

We also showed that TTV analyses of CoRoT data are promising for detecting
additional objects in transiting systems observed by the satellite.

%__________________________________________________ 
   \begin{table}
      \caption[]{Amplitudes and periods of $O-C$ variations in
      CoRoT-1b system.}
         \label{oc}
     	\begin{tabular}{llcc}
            \hline
            \noalign{\smallskip}
            Mass of  the &  Configuration & Amplitude & Period\footnote{In the units of consecutive transit numbers.}\\
            perturbing object &  &  in seconds & \\
            \noalign{\smallskip}
            \hline
            \noalign{\smallskip}
	    1 Earth mass & at $L_4$ point, $20^\circ$  & 60 & $\sim 10$\\
                         & libration amplitude      &    & \\
	    1 Earth mass & initially on circular orbit,  & 100 & $\sim 150$\\
	                 & 2:1 exterior mmr&    &      \\
	    30 Earth mass & outer planet initially  & 150 & $\sim 15$\\
	                  & with e=0.05 and & & \\
	                  & P=2.277218632 days & & \\
	    30 Earth mass & outer planet initially  & 150 & $\sim 30$\\  
	                  & with e=0.25 and & & \\
	                  & P=4.2679123 days & & \\
            \noalign{\smallskip}
            \hline
\end{tabular}
\end{table}

%\begin{figure*}%f1
%\includegraphics[width=6.9cm]{trojan1.eps}
%\caption{The numerically simulated $O-C$ diagram of CoRoT-1b system assuming
%a trojan system. In this configuration an Earth-mass planet librating with an
%amplitude of $20^\circ$ around L4 point.}
%\label{fig1}
%\end{figure*}

\begin{figure*}%f2
\includegraphics[width=6.9cm]{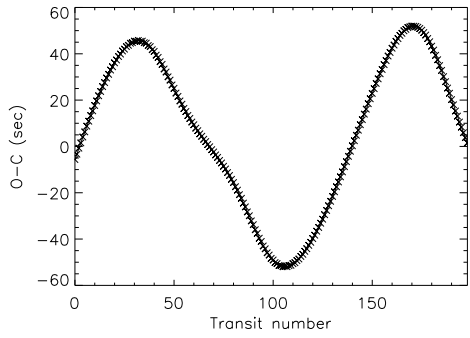}
\caption{The simulated O-C diagram of CoRoT-1b if the transiting planet is
perturbed by an Earth-mass planet initially on a circular orbit in 2:1 
mean motion resonance.}
\label{fig3}
\end{figure*}

\begin{figure*}%f5
\includegraphics[width=12.9cm]{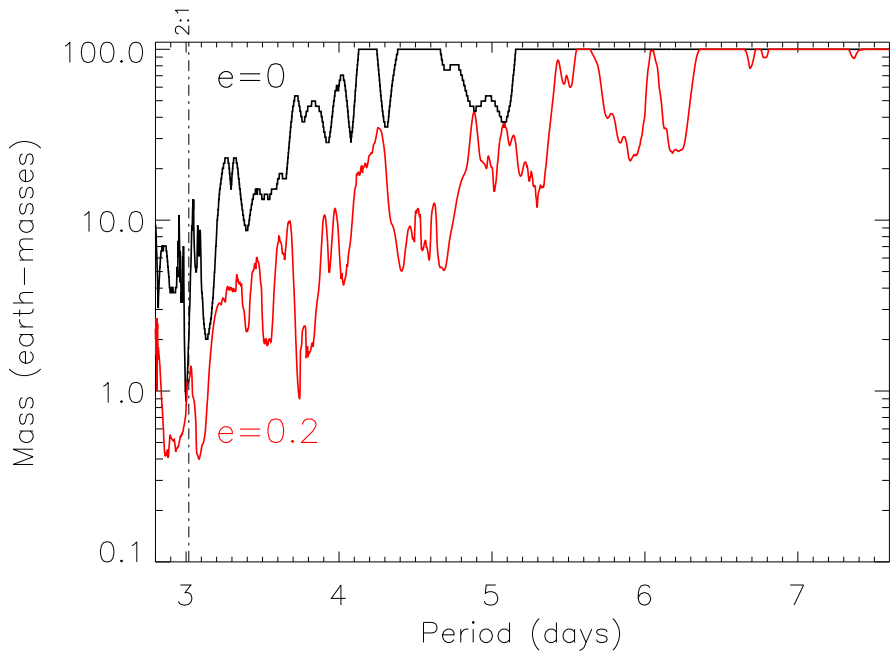}
\caption{Maximum allowed mass of a hypothetical perturbing object as a function of its
orbital period for excentricities e=0 and 0.2. The 2:1 mean
motion resonance is indicated.}
\label{fig3}
\end{figure*}

\begin{figure*}%f4
\includegraphics[width=8.9cm,angle=270]{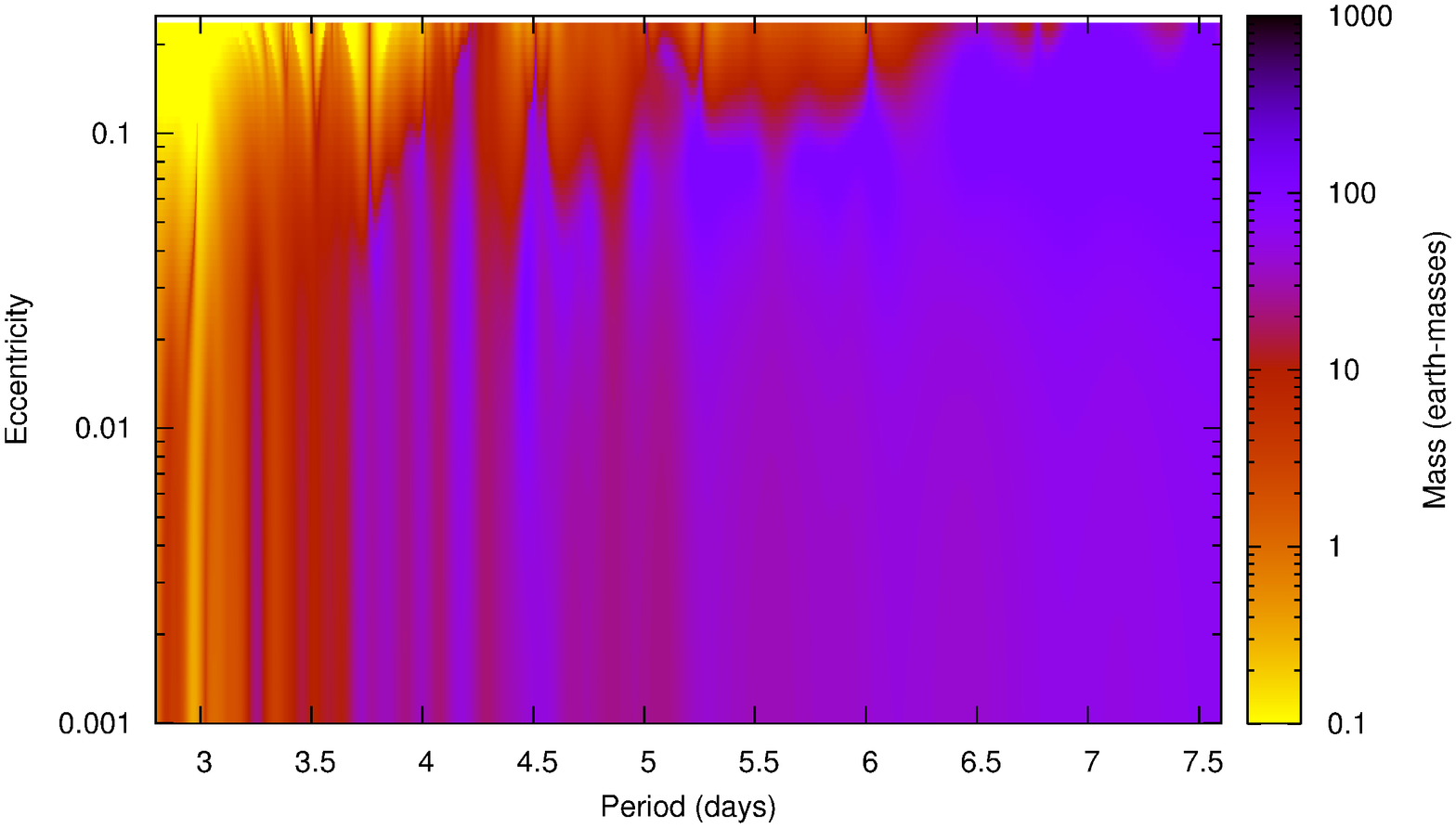}
\caption{Upper mass limits for a hypothetical second object in the CoRoT-1b system as
a function of the perturber's orbital period and eccentricity.}
\label{fig3}
\end{figure*}

%
%\begin{figure*}%f3
%\includegraphics[width=6.9cm]{eccentric1.eps}
%\caption{The same as Fig. 1 just here the perturbing object is a 30 earth-mass
%outer planet on an eccentric orbit (initially e=0.05 and P=2.77218632 days).}
%\label{fig3}
%\end{figure*}
%
%
%\begin{figure*}%f4
%\includegraphics[width=6.9cm]{compact1.eps}
%\caption{The same as Fig. 1 just here the perurbing object is a 30 earth-mass
%outer planet on an eccentric orbit (initially e=0.05 and P=4.26797123 days).}
%\label{fig4}
%\end{figure*}

\begin{acknowledgements}
The team at IAC acknowledges support by grant ESP2007-65480-C02-02 of
the Spanish Ministerio de Ciencia e Innovaci\'{o}n. The German CoRoT 
Team (TLS and Univ. Cologne) acknowledges DLR grants 50OW0204, 50OW0603,
and 50QP07011. EA thanks NSF for CAREER grant 0645416.
\end{acknowledgements}


\begin{thebibliography}{}

%   \bibitem[1980]{mizuno} Mizuno H. 1980,
%      Prog. Theor. Phys., 64, 544


\bibitem[2005]{agol05} Agol, E., Steffen, J., Sari, R., Clarkson, W. 2005,
        \mnras~359, 567

\bibitem[2007]{agol07} Agol, E., Steffen, J., 2007,
        \mnras~374, 941

\bibitem[2008]{alonsoetal} Alonso, R., Aigrain, S., Pont, F., Mazeh, T. et al.
	2008, arXiv0807.4828
	
\bibitem[2009]{auvergne} Auvergne, M., Bodin, P., Boisnard, L. et al. 2009, 
        A\&A, submitted

\bibitem[2007]{baglin} Baglin, A., Auvergne, M., Barge, P., et al. 2007, in American
Institute of Physics Conference Series, Vol. 895, American Institute of Physics
Conference Series, ed. C. Dumitrache, N. A. Popescu, M. D. Suran and V. Mioc,
201-209

\bibitem[2008]{barge} Barge, P., Baglin, A., Auvergne, M., Rauer, H., L\'eger, 
         A.. et al. 2008,
	 A\&A 482, 17 

\bibitem[2009]{bean} Bean, L. J., A\&A submitted, astro-ph/0903.1845 

\bibitem[2006]{Boisnard} Boisnard, L., Baglin, A., Auvergne, M., Deleuil, M.,
        Catala, C. 2006,
	in ESA Special Publication, Vol. 1306, 465-471

\bibitem[2003]{borko} Borkovits, T., \'Erdi, B., Forg\'acs-Dajka, E., Kov\'acs, T. 2003,
        A\&A 398, 1091	

\bibitem[1999]{chambers} Chambers, J. E. 1999
        \mnras~304, 793

\bibitem[2008]{coughlin} Coughlin, J. L., Stringfellow, G. S., Becker, A. C.; 
        López-Morales, M., Mezzalira, F., Krajci, T. 2008
        \apj~689, 149

\bibitem[2008]{diaz} D\'{\i}az, R. E. et al. 2008
        \apj~682, 49

\bibitem[2007]{ford} Ford, E. B., Holman, M. J. 2007,
        \apj~664, 51

\bibitem[2007]{heyl} Heyl, J. S., Gladman, B. J. 2007,
        \mnras~377, 1511

\bibitem[2007]{heyl} Holman, M. J., Murray, N. W. 2005,
        Science~307, 1288

\bibitem[2008]{hrudkova} Hrudkov\'a, M., Skillen, I., Benn, Ch., Pollacco, D.,
        Gibson, N., Joshi, Y., Harmanec, P., Tulloch, S. 2008, 	
	in: Transiting Planets, Proceedings of the International Astronomical 
	Union, IAU Symposium, Vol. 253, 446
	
%\bibitem[2002]{kalimeris} Kalimeris, A., Rovithis-Livaniou, H., Rovithis, P. 2002,
%        A\&A 387, 969

\bibitem[2009]{kipping} Kipping, D. M. 2009,
        \mnras~392, 181

%\bibitem[1956]{kwee} Kwee, K. K., van Woerden, H. 1956,
%        {\it Bull. Astron. Inst. Netherlands}, 12, 327

\bibitem[2005]{lenzbreger} Lenz, P., Breger, M., CoAst 146, 53

\bibitem[2002]{mandelagol} Mandel, K., Agol, E. 2002,
       \apj~580, 171

\bibitem[2007]{millerrricci} Miller-Ricci, E., Rowe, J. F., Sasselov, D., 
        Matthews, J. M., Guenther, D. B., Kuschnig, R., Moffat, A. F. J., 
	Rucinski, S. M., Walker, G. A. H., Weiss, W. W.	2007,
	ASPC 366, 146

\bibitem[2008a]{miller08a} Miller-Ricci, E., Rowe, J. F., Sasselov, D., 
        Matthews, J. M., Guenther, D. B., Kuschnig, R., Moffat, A. F. J., 
	Rucinski, S. M., Walker, G. A. H., Weiss, W. W. 2008a
	\apj~682, 586	

\bibitem[2008b]{miller08b} Miller-Ricci, E., Rowe, J. F., Sasselov, D., 
        Matthews, J. M., Kuschnig, R., Croll, B., Guenther, D. B., Moffat, 
	A. F. J., Rucinski, S. M., Walker, G. A. H., Weiss, W. W. 2008b
	\apj~682, 593	

\bibitem[2002]{miralda}	Miralda-Escud\'e, J. 2002,
        \apj~564, 1019 

\bibitem[2008]{nesvorny} Nesvorn\'y, D., Morbidelli, A. 2008,
        \apj~688, 636
	
\bibitem[2008]{palandras} P\'al A., Kocsis B. 2008,
        \mnras~389, 191 

%\bibitem[2008]{pejcha} Pejcha, O. 2008, 'Fit Procedure Description', available
%	at: {\it http://var2.astro.cz/ETD/}

\bibitem[2007]{pont} Pont, F., Gilliland, R. L., Moutou, C., Charbonneau, D.,
        Bouchy, F., Brown, T. M., Mayor, M., Queloz, D., Santos, N., Udry, S.	
        2007, A\&A 476, 1347

\bibitem[1992]{press} Press et al. 1992, {\it Numerical recipes}, Cambridge
        University Press

\bibitem[2009]{rabus} Rabus, M., Deeg, H. J., Alonso, R., Belmonte, J. A.,
         Almenara, J. M., 2009, A\&A, accepted

\bibitem[1994]{schneider94} Schneider, J. 1994,  P\&SS 42, 539

\bibitem[2004]{schneider04} Schneider, J. 2004,  ESASP 538, 407

\bibitem[2007]{simon} Simon, A., Szatm\'ary, K., Szab\'o, Gy. M. 2007,
        A\&A 470, 727

\bibitem[2005]{steffen05} Steffen, J. H., Agol, E. 2005
       \mnras~364, 96

\bibitem[2006]{steffen06} Steffen, J. H. 2006
	{\it PhD thesis, University of Washington}

\bibitem[2007]{steffen07} Steffen, J. H., Gaudi, B. S., Ford, E. B., Agol, E.,
        Holman, M. J. 2007, arXiv0704.0632

\bibitem[2009]{stringfellow} Stringfellow, G. S., Coughlin, J. L., 
        L\'opez-Morales, M., Becker, A. C., Krajci, T., Mezzalira, F., Agol, E.
	2009, in: Cool Stars, Stellar Systems and the Sun,Proceedings of the 
	15th Cambridge Workshop on Cool Stars, Stellar Systems and the Sun. 
	AIP Conference Proceedings, Vol. 1094, 481
	 
\bibitem[2007]{valencia07}  Valencia, D., Sasselov D. D., O'Connell, R. J.,
        2007,
	\apj~656, 545

\end{thebibliography}
\end{document}